\documentclass{jpp}

\usepackage{graphicx}
\usepackage{natbib}

\ifCUPmtlplainloaded \else
  \checkfont{eurm10}
  \iffontfound
    \IfFileExists{upmath.sty}
      {\typeout{^^JFound AMS Euler Roman fonts on the system,
                   using the 'upmath' package.^^J}%
       \usepackage{upmath}}
      {\typeout{^^JFound AMS Euler Roman fonts on the system, but you
                   dont seem to have the}%
       \typeout{'upmath' package installed. JPP.cls can take advantage
                 of these fonts, if you use 'upmath' package.^^J}%
      }
  \else
  \fi
\fi

\ifCUPmtlplainloaded \else
  \checkfont{msam10}
  \iffontfound
    \IfFileExists{amssymb.sty}
      {\typeout{^^JFound AMS Symbol fonts on the system, using the
                'amssymb' package.^^J}%
       \usepackage{amssymb}%
         
         \let\geq=\geqslant
      }{}
  \fi
\fi

\ifCUPmtlplainloaded \else
  \IfFileExists{amsbsy.sty}
    {\typeout{^^JFound the 'amsbsy' package on the system, using it.^^J}%
     \usepackage{amsbsy}}
    {}
\fi

\newsavebox{\astrutbox}
\sbox{\astrutbox}{\rule[-5pt]{0pt}{20pt}}

\title[Particles trajectories in Weibel magnetic filaments with a flow-aligned magnetic field]{Particles trajectories in Weibel magnetic filaments with a flow-aligned magnetic field}

\author[A. Bret]%
{A\ls N\ls T\ls O\ls I\ls N\ls E\ns B\ls R\ls E\ls T $^{1,2}$%
  \thanks{Email address for correspondence: antoineclaude.bret@uclm.es}
}

\affiliation{$^1$ETSI Industriales, Universidad de Castilla-La Mancha, 13071 Ciudad Real, Spain\\[\affilskip]
$^2$Instituto de Investigaciones Energ\'{e}ticas y Aplicaciones Industriales, Campus Universitario de Ciudad Real, 13071 Ciudad Real, Spain}

\date{?; revised ?; accepted ?. - To be entered by editorial office}
\begin{document}

\maketitle

\begin{abstract}
For a Weibel shock to form, two plasma shells have to collide and trigger the Weibel instability. At saturation, this instability generates in the overlapping region magnetic filaments with peak field $B_f$. In the absence of an external guiding magnetic field, these filaments can block the incoming flow, initiating the shock formation, if their size is larger than the Larmor radius of the incoming particles in the peak field. Here we show that this results still holds in the presence of an external magnetic field $B_0$, provided it is not too high. Yet, for $B_0 \gtrsim B_f/2$, the filaments become unable to stop any particle, regardless of its initially velocity.
\end{abstract}


\section{Introduction}
Magnetic filaments are spontaneously generated by the growth of the Weibel, or filamentation, instability. This instability is triggered when two relativistic plasma shells cross each other \cite{Huntington2015}. It has been invoked for intergalactic magnetic fields generation \cite{Schlickeiser2003}, inertial confinement fusion \cite{Silva2002,DeutschPRE2005}, or collisionless shocks physics \cite{Medvedev1999,BretPoP2013,Fiuza2012}. Regarding collisionless shocks, it has been known for long that they are capable of accelerating particles \cite{Blandford1987} (for a particle to receive energy, and keep it, the medium has to be collisionless). As such, it is believed that they may play a key-role in the generation of high energy cosmic rays, or gamma ray bursts \cite{Vietri2003ApJ,Piran2005}.

During the last decade or so, the physics of collisionless shocks has undergone a renewed interest, as it became possible to study them in detail through large-scale particle-in-cell simulations \cite{SilvaApJ,Spitkovsky2005,niemiec2012}. On the theory side, one of the problems currently being solved has to do with the very formation of such shocks. Concerning fluid shocks, it is known they can arise from the steepening of a large amplitude sound wave, or from the collision of two media, with a collision speed faster than the speed of sound in one of them \cite{Zeldovich}. The formation process in the collisionless case is more involved because such shocks occur in collionless plasmas. In these kind of settings, where the collision frequency is virtually zero, a shock has to be mediated by collective plasma interactions \cite{Sagdeev66}.

As electromagnetic objects, collisionless shocks display far more variety than fluid shocks. They can be electrostatic or electromagnetic (Weibel) \cite{Sarri2011,Dieckmann2014NJP,Stockem2014}, form in pair or electron/ion plasmas \cite{BretPoP2013,BretPoP2014,Bret2015ApJL}, and on the top of these dichotomies, be influenced by the strength of an external magnetic field and its orientation \cite{Treumann2009,Marco2016}.

Among the different kinds of collisionless shocks, an interesting sub-class is formed by the so-called ``Weibel shocks''. When two collisionless plasma shells run into each other, the overlapping region turns unstable. If the encounter occurs at relativistic velocity, the dominant instability is the Weibel one \cite{BretPoPHierarchie,BretPRL2008,BretPoPReview}. This instability grows magnetic filaments which can block the incoming flow, initiating the shock formation \cite{BretPoP2013,BretPoP2014,Bret2015ApJL}.

Note that at the very beginning of the shock formation process, the counter-streaming plasmas cross each other, triggering the Weibel instability. At this stage, magnetic filaments are under formation, and particles are not stopped. Then, the Weibel instability reaches saturation, the filaments are fully formed, and may stop the particles arriving at later times. The flow stopped by the filaments is not the flow that formed them.

It is clear that in shock setting, once the filaments have stopped a number of particles, these may perturb the filaments and break them. Yet, it has been found that the resulting field configuration is at least as efficient as were the filaments to stop the flow \cite{BretPoP2014,Bret2015ApJL}. In order to know whether a shock will start forming after the saturation of the Weibel instability, or not, the key question is therefore: at saturation time, are the filaments able to block the incoming particles?

For the case where there is no external magnetic field, the conditions upon which the flow is blocked can been derived from the analysis of the motion of charged particles in the Weibel magnetic filaments \cite{BretPoP2015}. Yet, many astrophysical settings are magnetized. If two plasma shells interact over an external, flow-aligned, magnetic field $\mathbf{B}_0$, the Weibel instability still grows while the field is not too strong \cite{Godfrey1975,BretPoPMagne}. Here, the resulting magnetic filaments supposed to block the incoming flow for the shock to form, will be superimposed over the external $\mathbf{B}_0$. The goal of this article is therefore to study this problem: how about the trajectory of charged particles within magnetic filaments and an external $\mathbf{B}_0$?

\section{Model considered}
\begin{figure}
\includegraphics[width=0.6\textwidth]{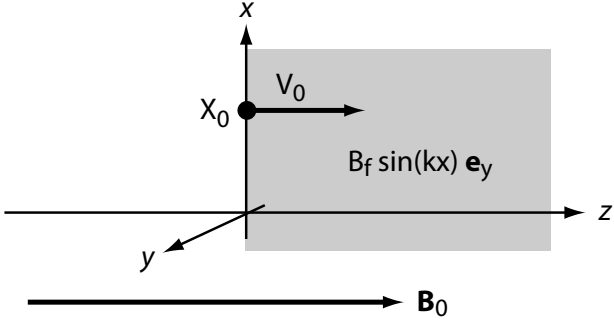}
\caption{Setup considered. The field $B_f\sin kx \mathbf{e}_y$ fills the half-space $z>0$ (shaded area). The field $B_0 \mathbf{e}_z$ fills the entire space. A charged particle starts from $\mathbf{x}=(x_0,0,0)$ at $t=0$, with $\dot{\mathbf{x}}=(0,0,v_0)$.}
\label{setup}
\end{figure}

We therefore consider the setup pictured on Figure \ref{setup}. The half space $z>0$ is filled with the magnetic filaments $\mathbf{B}_f=B_f\sin kx \mathbf{e}_y$, where $k^{-1}$ is therefore the coherence length of the magnetic field. A static, homogenous, external magnetic field  $\mathbf{B}_0=B_0 \mathbf{e}_z$ fills the entire space. A particle with charge $q$, mass $m$, velocity $\mathbf{v}_0=v_0\mathbf{e}_z$ and Lorentz factor $\gamma=(1-v_0^2/c^2)^{-1/2}$, is launched at $t=0$ from $x=x_0$ and $y=z=0$.

Note that such a Heaviside modelling already represents a significant simplification of the interface \cite{BretPoP2015}. The electromagnetic structures that have been identified within the finite width interface \cite{Milosavljevic2006,Pathak2015} cannot be accounted for in the present work. However, previous analysis of the shock formation process have shown that the particles are indeed stopped within the Weibel filaments \cite{lyubarsky06,BretPoP2014,Bret2015ApJL}.

Having only the Lorentz force involved, the Lorentz factor is constant, and the equation of motion reads,

\begin{equation}\label{eq:motion}
  m\gamma \ddot{\mathbf{x}} = q\frac{\dot{\mathbf{x}}}{c}\times (\mathbf{B}_f + \mathbf{B}_0).
\end{equation}

With the aforementioned choice of the fields, this equation splits into the 3 following scalar ones,
\begin{eqnarray}
  m\gamma\ddot{x} &=& -   \frac{q}{c}( \dot{z}B_f\sin kx - \dot{y}B_0),  \nonumber \\
  m\gamma\ddot{y} &=& -   \frac{q}{c}\dot{x}B_0,  \nonumber \\
  m\gamma\ddot{z} &=&     \frac{q}{c} \dot{x}B_f \sin kx. \nonumber
\end{eqnarray}

We now proceed to the following change of variables,
\begin{equation}\label{eq:change}
\mathbf{x} \rightarrow \mathbf{X}/k,~~B_0 \rightarrow \alpha B_f, ~~ t \rightarrow \tau/\omega_B,~~\mathrm{with}  ~~ \omega_B = \frac{qB_f}{\gamma mc},
\end{equation}
and obtain the following equations,
\begin{eqnarray}\label{eq:dimless}
\ddot{X} &=& -   \dot{Z} \sin X + \alpha \dot{Y},  \nonumber \\
\ddot{Y} &=& -   \alpha \dot{X},   \\
\ddot{Z} &=&     \dot{X} \sin X, \nonumber
\end{eqnarray}
with initial conditions,
\begin{eqnarray}
\mathbf{X}(\tau=0) &=& \left( kx_0,0,0 \right)  \nonumber \\
\dot{\mathbf{X}}(\tau=0) &=& \left(0,0,\frac{kv_0}{\omega_B} \right).
\end{eqnarray}

\begin{figure}
\includegraphics[width=0.6\textwidth]{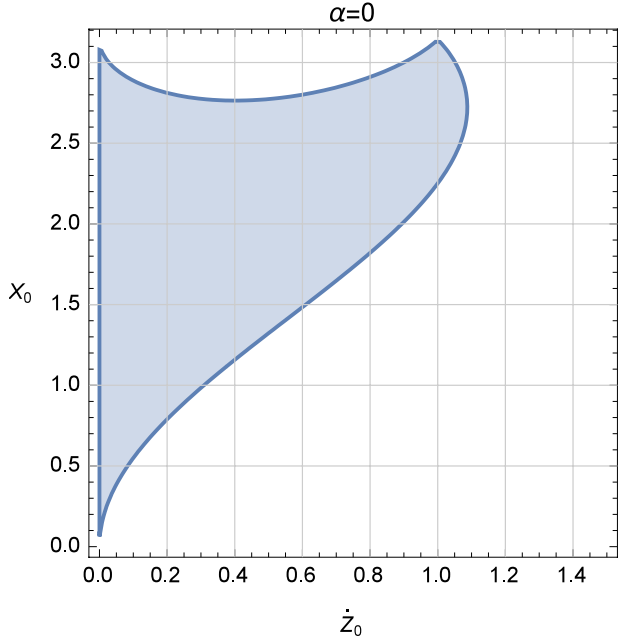}
\caption{Case with no external $\mathbf{B}_0$ ($\alpha=0$) \cite{BretPoP2015}. Whenever the initial conditions pertain to the shaded area, the particles comes back to $z<0$. A particle launched from $X_0=\pi$ cannot bounce back since $\mathbf{B}_f(x=\pi)=0$.}
\label{zero}
\end{figure}

The case $\alpha=0$ as already been treated \cite{BretPoP2015}. The domain of the phase space $(\dot{Z}_0,X_0)$ where the particle bounces back against the region $z>0$ is pictured on Figure \ref{zero}. Whenever the initial conditions pertain to the shaded area, the particles comes back to $z<0$.

In collisionless shocks conditions, the interface at $z\sim 0$ is the location of important magnetic turbulence. As a consequence, particles starting from the shaded area do not systematically bounce back to $z<0$, but tend to be trapped instead, in the region $z>0$. In contrast, particles starting out of the shading area stream through the magnetic filaments.

Our goal from now-on is to find out how Figure \ref{zero} evolves when the external magnetic field $\mathbf{B}_0$ is switched on, that is, when $\alpha \neq 0$.

\section{Analytical results}
\subsection{Symmetries}
Although the full resolution of the problem will be numerical, a number of analytical considerations can pave the way to such resolution.

To start with, the system (\ref{eq:dimless}) is invariant by the substitutions $\alpha \rightarrow -\alpha$ and $X \rightarrow -X$. It means an opposite $\mathbf{B}_0$ will simply have the particles spin in the opposite direction. But the $Z$ component, which determines whether the particle bounces back or not, is left unchanged. We can therefore restrict our study to $\alpha>0$, that is, $B_0>0$.

Also, we can integrate the third equation of the system (\ref{eq:dimless}), and obtain regardless of $\alpha$,
\begin{equation}\label{eq:bounded}
  \dot{Z}(\tau) - \dot{Z}_0 = \cos X_0 - \cos X(\tau).
\end{equation}
Since the velocity is constant, equal to $|\dot{\mathbf{X}}(\tau=0)|=\dot{Z}_0$, the left-hand-side remains negative along the trajectory. As a consequence, $\cos X(\tau)  \geq \cos X_0$, $\forall \tau > 0$. Therefore, trajectories are bounded within intervals of width $2\pi$ and one can restrict the study to $X \in [-\pi, \pi]$ \cite{BretPoP2015}.

Finally, the system (\ref{eq:dimless}) is also invariant when substituting $X \rightarrow -X$ and $Y \rightarrow -Y$. This implies that we can restrict our attention to the interval $X \in [0, \pi]$.

\subsection{Equation for $X$ for $Z>0$}
The second equation of the system (\ref{eq:dimless}) can be integrated, giving,
\begin{equation}\label{eq:Y}
 \dot{Y} = -   \alpha (X-X_0).
\end{equation}
Inserting this result in the first equation of the system, and expressing $\dot{Z}$ in terms of $X$ with Eq. (\ref{eq:bounded}), we obtain the following equation for $X$ only,
\begin{equation}\label{eq:X}
 \ddot{X} +   \sin X (\cos X_0 +\dot{Z}_0 - \cos X) + \alpha^2 (X-X_0) = 0.
\end{equation}
While it cannot be solved exactly, an approximated resolution is possible when $\sin X \sim 0$.

\subsection{Motion near $X = 0$}
For $X_0 \sim 0$, we can write $\sin X \sim X$ and check later that $X(\tau)$ remains close to 0 for $\tau>0$. Equation (\ref{eq:X}) becomes after linearization,
\begin{equation}\label{eq:Xapprox}
 \ddot{X} +   \omega^2 X   = \alpha^2 X_0~~\mathrm{with}~~\omega^2 = \alpha^2 + \dot{Z}_0 > 0.
\end{equation}
The solution reads,
\begin{equation}\label{eq:Xapprox_sol}
 X(\tau) = \frac{\alpha ^2 X_0  }{\omega^2} + \frac{\dot{Z}_0 X_0 }{\omega^2}\cosh(i \omega \tau)~~\mathrm{with}~~i^2=-1.
\end{equation}
With $\omega^2 > 0$, trajectories oscillate between $X_0$ and $-X_0(1-2\alpha^2/\omega^2)$, remaining thus confined around $X=0$ since $\alpha^2 < \omega^2$.

We can then integrate the third equation of the system (\ref{eq:dimless}) to obtain an explicit expression for $Z(\tau)$. The result is,
\begin{equation}\label{eq:Zapprox}
 Z(\tau) = \lambda \tau - \frac{\dot{Z}_0 X_0^2}{4 (\dot{Z}_0 + \alpha^2)^{5/2}}
 \left[4 \alpha^2 + \dot{Z}_0 \cosh (i \omega \tau)\right] \sinh (i \omega \tau),
\end{equation}
with
\begin{equation}
\lambda = \frac{\dot{Z}_0}{4} \left[4-\frac{X_0^2 (4 \alpha ^2+\dot{Z}_0)}
{(\alpha ^2+\dot{Z}_0)^2}\right].
\end{equation}

We may now find a condition for the particle to return to the region $Z<0$, or to stream through the region $Z>0$. The second term of Eq. (\ref{eq:Zapprox}) is oscillatory. But the first one, namely $\lambda \tau$, tends to $\pm \infty$ depending on the sign of the parameter $\lambda$. We can indeed write, $\lim_{\tau +\infty} Z(\tau) = \mathrm{sign} (\lambda)\infty$. Therefore, our particle returns to $Z<0$ if $\lambda < 0$, that is,
\begin{equation}
\frac{(\alpha ^2+\dot{Z}_0)^2}{4 \alpha ^2+\dot{Z}_0} < \frac{X_0^2}{4}.
\end{equation}
This expression bears interesting consequences. In the absence of external magnetic field, that is, $\alpha=0$, the equation above gives $\dot{Z}_0 < X_0^2/4$ \cite{BretPoP2015}. As can be seen on Figure \ref{zero}, the frontier between particles with $Z(\infty)=\pm\infty$ goes all the way down to $(\dot{Z}_0,X_0)=(0,0)$.

Now, setting $\alpha\neq 0$ and $\dot{Z}_0 =0$ in the equation above gives,
\begin{equation}
- \alpha < X_0 <  \alpha.
\end{equation}
(Note that this conclusion reflects the $\pm X$ symmetry already mentioned.) As a consequence, particles starting from $X_0 < \alpha$ stream through the magnetized filaments, regardless of their initial velocity $\dot{Z}_0$.

\subsection{Motion near $X = \pi$}
For $X_0 \sim \pi$, we can write $\sin X \sim \pi - X$ and again, check afterward whether the trajectory remains confined near $X=\pi$ or not.

Setting $\tilde{X} = X-\pi$ and $\tilde{X}_0 = X_0-\pi$, and linearizing equation (\ref{eq:X}), we now obtain,
\begin{equation}\label{eq:XapproxPi}
 \ddot{\tilde{X}} +   \omega^2 \tilde{X}   = \alpha^2 \tilde{X}_0~~\mathrm{with}~~\omega^2 = \alpha^2 - \dot{Z}_0,
\end{equation}
which solution reads,
\begin{equation}\label{eq:Xapprox_solPi}
 \tilde{X}(\tau) = \frac{\alpha ^2 \tilde{X}_0  }{\omega^2} - \frac{\dot{Z}_0 \tilde{X}_0 }{\omega^2}\cosh(i \omega \tau).
\end{equation}
We need now discuss the two cases corresponding to the sign of $\omega^2$.
\begin{itemize}
  \item If $\omega^2  < 0$, that is, $\alpha^2 < \dot{Z}_0$, $\omega$ is purely imaginary and  trajectories diverge on a time scale $\tau \sim |\omega|^{-1}$, rendering expression (\ref{eq:Xapprox_solPi}) inaccurate. Note that $B_0=0$ implies $\alpha=0$, which means that in the absence of a guiding field, trajectories cannot remain confined \cite{BretPoP2015}.
  \item If $\omega^2  > 0$, that is, $\alpha^2 > \dot{Z}_0$, trajectories now oscillate between $\tilde{X}_0$ and $\tilde{X}_0^+=\tilde{X}_0(1+2\alpha^2/\omega^2)$.  Yet, in the present case where $\alpha^2 > \dot{Z}_0$
      we have $\alpha^2/\omega^2 > 1$ so that $\tilde{X}_0^+$ may not remain close to zero.
      But if $\tilde{X}_0^+ \ll 1$, the approximate solution (\ref{eq:Xapprox_sol}) holds for any $\tau > 0$. We can linearize the third equation of the system (\ref{eq:dimless}) near $X=\pi$, obtaining for $\tilde{X}(\tau)$,
\begin{equation}
 \ddot{Z} = -\dot{\tilde{X}}\tilde{X}.
\end{equation}
The solution can again be written as the sum of an oscillatory term, plus a term $\lambda \tau$, with now
\begin{equation}
\lambda = \frac{\dot{Z}_0}{4} \left[4-\frac{\tilde{X}_0^2 (4 \alpha ^2-\dot{Z}_0)}
{(\alpha ^2-\dot{Z}_0)^2}\right].
\end{equation}
As was the case for $X_0 \sim 0$, particles stream through the magnetized region if $\lambda > 0$. For $\dot{Z}_0 \rightarrow 0^+$, this implies
\begin{equation}
\tilde{X}_0^2 < \alpha^2 \Rightarrow \pi - \alpha < X_0 < \pi + \alpha.
\end{equation}
Particles meeting the condition above  will stream through the $Z>0$ region (note that the condition $\alpha^2 > \dot{Z}_0$, necessary to have $\omega^2 > 0$, is implicitly met for $\dot{Z}_0 \rightarrow 0^+$).
\end{itemize}

The forthcoming numerical study will confirm the conclusions of these two subsections, as evidenced on Figures \ref{alpha}.

\subsection{Canonical momenta}
We here check that the constancy of 2 canonical momenta and of the Hamiltonian does not bring new information capable of pushing further our analytical exploration. The total magnetic field in the region $z>0$ reads $\mathbf{B} = (0, B_f \sin kx, B_0)$. It can be derived from the vector potential,
\begin{equation}
\mathbf{A} = \left(0, B_0 x, B_f \frac{\cos kx}{k} \right), ~~\mathrm{with}~~ \nabla \times \mathbf{A} = \mathbf{B}.
\end{equation}
The canonical momentum of a particle of mass $m$ and charge $q$ in the $z>$ region reads \cite{jackson1998},
\begin{equation}
\mathbf{P} = \mathbf{p} + \frac{q}{c} \mathbf{A},
\end{equation}
with $\mathbf{p}=\gamma m \mathbf{v}$. The canonical momentum $\mathbf{P}$ therefore only depends on the time derivative of the coordinates, and on the $x$ coordinate. It follows that the $y$ and $z$ coordinates of $\mathbf{P}$ are constants of the motion. Using the dimensionless variables (\ref{eq:change}), this reads
\begin{eqnarray}
\mathbf{P}_y &=& \mathbf{P}_y(t=0) ~~ \Rightarrow ~~ \alpha X + \dot{Y} = \alpha X_0, \\
\mathbf{P}_z &=& \mathbf{P}_z(t=0) ~~ \Rightarrow ~~ \dot{Z} + \cos X = \dot{Z}_0 + \cos X_0. \nonumber
\end{eqnarray}
The first equation of the system above is identical to Eq. (\ref{eq:Y}), while the second is identical to Eq. (\ref{eq:bounded}). Furthermore, the constancy of the Hamiltonian $\mathcal{H}= c \sqrt {m^2 c^2 + {\left( c\mathbf{P} - q \mathbf{A}  \right) }^2}$, reduces to $\mathbf{p}^2=$cst, which has already been accounted for through $\gamma=$cst.

\begin{figure}
\includegraphics[width=0.6\textwidth]{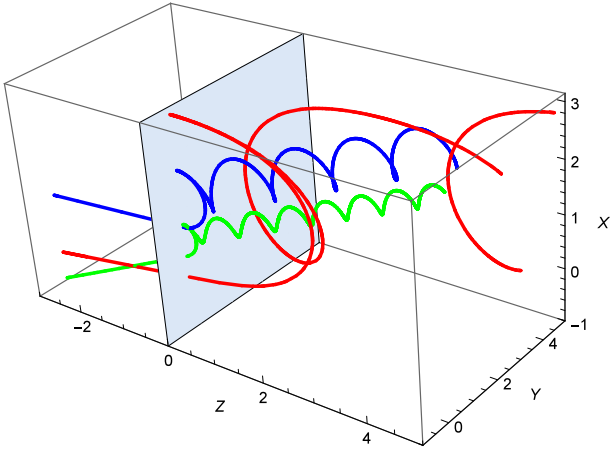}
\caption{Numerical computation of a few trajectories. Red: $X_0=3$, $\dot{Z}_0=1$ and $\alpha=0$ or 0.2. Blue:  $X_0=2$, $\dot{Z}_0=0.5$ and $\alpha=0$ or 0.5. Green: $X_0=1$, $\dot{Z}_0=0.2$ and $\alpha=0$ or 0.5. In each case, the trajectory with $\alpha \neq 0$ goes to $Z=+\infty$. The gray area features the plane $Z=0$.}
\label{traj}
\end{figure}

\section{Numerical study}
Trajectories can be studied numerically using the \textbf{NDSolve} function of the \emph{Mathematica} software. A series of 3D trajectories are pictured on Figure \ref{traj}. The 2 red trajectories both have $X_0=3$ and $\dot{Z}_0=1$. But $\alpha=0$ or 0.2 makes one bouncing back, and the other going to $Z=+\infty$. The 2 blue trajectories both have $X_0=2$ and $\dot{Z}_0=0.5$. But $\alpha=0$ or 0.5 makes one bouncing back, and the other going to $Z=+\infty$. Finally, the 2 green trajectories both have $X_0=1$ and $\dot{Z}_0=0.2$. But $\alpha=0$ or 0.5 makes one bouncing back, and the other going to $Z=+\infty$.

\begin{figure}
\includegraphics[width=\textwidth]{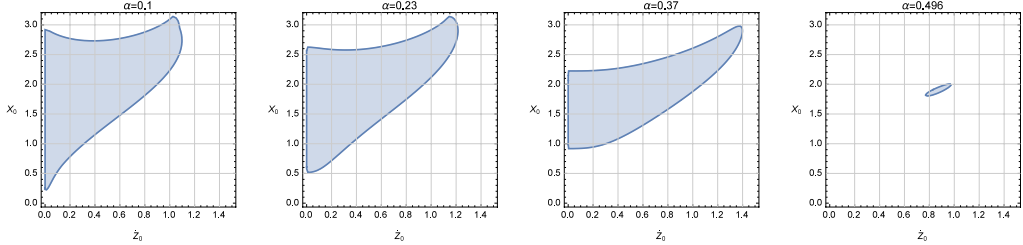}
\caption{Evolution of Figure \ref{zero} for $\alpha$ varying up to $\sim 0.5$. Particles starting from the shaded area go back to $Z<0$. This domain vanishes for $\alpha > 0.497$, around $(\dot{Z}_0,X_0)\sim (0.87,1.9)$.}
\label{alpha}
\end{figure}

The evolution of Figure \ref{zero} for $\alpha$ varying up to $\sim 0.5$ is pictured on Figure \ref{alpha}. The expected shrinking of the shaded area at $\dot{Z}_0=0$ is observed, while a non-trivial reduction is also observed for  $\dot{Z}_0>0$. The overall picture makes perfect sense: as its relative intensity increases, the guiding magnetic field forces more and more particles to follow trajectories leading to $Z=+\infty$.

\begin{figure}
\includegraphics[width=0.6\textwidth]{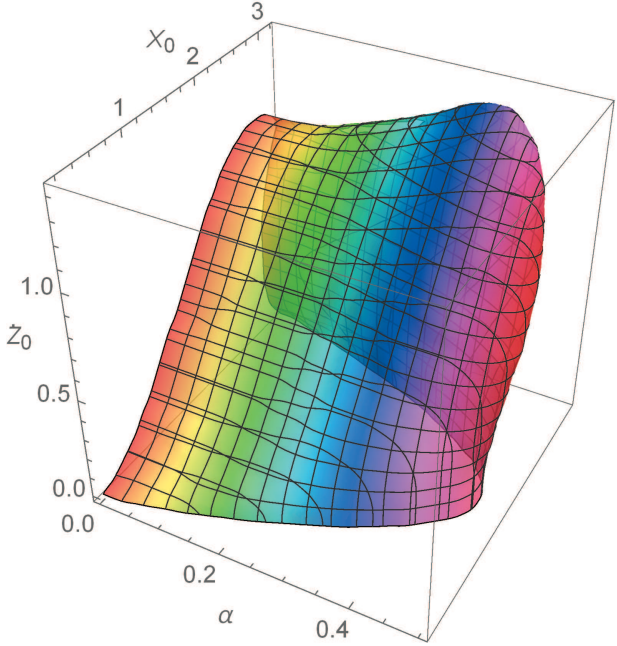}
\caption{3D representation of the shaded volume in the $(\dot{Z}_0,X_0,\alpha)$ space, resulting from the assembling of the shaded areas at $\alpha=$cst, like the ones pictured on Fig. \ref{alpha}. Parameters located inside the volume yield particle trajectories ultimately bouncing back to $Z<0$. The color code refers to the slices $\alpha=$ cst.}
\label{3d}
\end{figure}

The 3D Figure \ref{3d} allows for a unified perception of the cuts at $\alpha=$cst represented on Fig \ref{alpha}.   One of the most remarkable features of this volume is that it is bounded. Whenever $\alpha > 1/2$ (rounding-up 0.497 to 1/2), particles end-up at $Z=+\infty$, regardless of their initial conditions $(\dot{Z}_0,X_0)$. At such a level of external magnetization, $\mathbf{B}_0$ systematically overcomes the effect of the magnetic filaments. We shall now check the expected consequences on the formation of a parallel Weibel shock in pair plasmas.

\section{Consequences for the formation of a parallel Weibel shock in pair plasmas}
Assume two identical cold plasma shells with density $n_0$, velocity $\pm \mathbf{v}_0$ and Lorentz factor $\gamma_0$ heading towards each other. The system is embedded in an external magnetic field $\mathbf{B}_0 \parallel \mathbf{v}_0$. As they overlap, we assume the interaction is mediated by the Weibel instability, with growth rate \cite{Stockem2006ApJ},
\begin{equation}\label{eq:gr}
  \frac{\delta}{\omega_p} =\sqrt{\delta_0^2 - \Omega_{B_0}^2},~~\mathrm{where}~~\Omega_{B_0} =\frac{1}{\omega_p} \frac{q B_0}{\gamma m c},
\end{equation}
and $\delta_0 = \beta \sqrt{2/\gamma}$ is the growth rate, in $\omega_p$ units, when $B_0=0$. The Weibel instability will generate magnetic filaments with a peak field $B_f$ at saturation fulfilling \cite{davidsonPIC1972,BretPoP2013},
\begin{equation}
  \frac{q B_f}{\gamma m c} = \delta \Leftrightarrow \delta_0^2 = \Omega_{B_0}^2 + \Omega_{B_f}^2,~~\mathrm{where}~~\Omega_{B_f} =\frac{1}{\omega_p} \frac{q B_f}{\gamma m c}.
\end{equation}
Writing now $B_f = B_0/\alpha$, we find that $\alpha < 1/2$ is equivalent to,
\begin{eqnarray}
  \alpha^2 = \frac{1}{(\delta_0/\Omega_{B_0})^2-1} &>& \frac{1}{4} \nonumber \\
  \Rightarrow  \sigma &>& \frac{2}{5},
\end{eqnarray}
where (we set $\beta=1$),
\begin{equation}
   \sigma = \frac{B_0^2/4\pi}{\gamma n_0 m c^2},
\end{equation}
is the well-known magnetization parameter. For practical purposes, this parameter is usually much smaller than 1 in astrophysical environments, even in such strong magnetized environments like pulsar wind nebulae \cite{Marco2016}. Even $\sigma=10^{-1}$ results in $\alpha=0.1$, with effects hardly distinguishable from $\alpha=0$.

\begin{figure}
\includegraphics[width=0.6\textwidth]{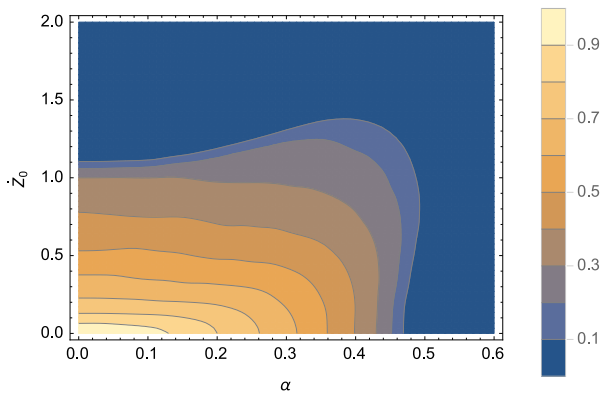}
\caption{Amount of each plasma shell bouncing back against the magnetic filaments region, as a function of $\dot{Z}_0$ and $\alpha$.}
\label{bounce}
\end{figure}

Since the two plasma shells are cold, the percentage of each shell which bounces back is simply obtained averaging our results over $X_0$. The result of the calculation is displayed on Figure \ref{bounce}, in terms of  $\dot{Z}_0$ and $\alpha$. As long as $\alpha \lesssim 0.4$, the beam goes through the magnetic filaments region for $\dot{Z}_0 > 1$, that is, $k^{-1} < v_0/\omega_B$. In other words, the beam streams through the magnetic filaments if their size is smaller than the Larmor radius of the particles in the peak field $B_f$. Beyond $\alpha > 1/2$, the totality of the beam always goes to $z=+\infty$.

\section{Conclusion}
The formation of a Weibel shock involves to collision of two plasma shells. As they pass through each other, the overlapping region becomes Weibel unstable. At saturation, it forms magnetic filaments of transverse size $1/k$ and peak field $B_f$. We here determined under which conditions these filaments block the incoming flow, thus initiating the shock formation.

It has already been found than in the absence of an external magnetic field, most of the flow (if cold) keeps streaming through the filaments, or not, whether $1/k$ is smaller or larger than the Larmor radius of the particles in the field $B_f$.

In the presence of a parallel magnetic field $\mathbf{B}_0$, everything relies on the parameter $\alpha=B_0/B_f$. While $\alpha \lesssim 0.4$, the overall picture is similar to that with $\alpha =0$. But for $\alpha>1/2$ a transition occurs. The guiding field $\mathbf{B}_0$ becomes dominant and particles keep streaming through the filaments regardless of their initial velocities. In realistic scenarios however, the parameter $\sigma$ which governs the transition, forces $\alpha \ll 1$.

Further works could contemplate the case of a perpendicular field. Such a scenario involves more free parameters than the current one since the field will have to be specified by its two normal components, instead of one single parallel component in the present case.

\section{Acknowledgments}
This work was supported by grant ENE2013-45661-C2-1-P from the Ministerio de Educaci\'{o}n y Ciencia, Spain
and grant PEII-2014-008-P from the Junta de Comunidades de Castilla-La Mancha.


\begin{thebibliography}{34}
\expandafter\ifx\csname natexlab\endcsname\relax\def\natexlab#1{#1}\fi

\bibitem[Blandford \& Eichler(1987)]{Blandford1987}
{\sc Blandford, R \& Eichler, D} 1987 Particle acceleration at astrophysical
  shocks: a theory of cosmic ray origin. {\em Phys. Rep.\/} {\bf 154}, 1.

\bibitem[Bret(2015)]{BretPoP2015}
{\sc Bret, A.} 2015 Particles trajectories in magnetic filaments. {\em Physics
  of Plasmas\/} {\bf 22}.

\bibitem[Bret \& Deutsch(2005)]{BretPoPHierarchie}
{\sc Bret, A. \& Deutsch, C.} 2005 Hierarchy of beam plasma instabilities up to
  high beam densities for fast ignition scenario. {\em Physics of Plasmas\/}
  {\bf 12}, 082704.

\bibitem[Bret {\em et~al.\/}(2006)Bret, Dieckmann \& Deutsch]{BretPoPMagne}
{\sc Bret, A., Dieckmann, M. \& Deutsch, C.} 2006 Oblique electromagnetic
  instabilities for a hot relativistic beam interacting with a hot and
  magnetized plasma. {\em Phys. Plasmas\/} {\bf 13}, 082109.

\bibitem[Bret {\em et~al.\/}(2008)Bret, Gremillet, B\'{e}nisti \&
  Lefebvre]{BretPRL2008}
{\sc Bret, A., Gremillet, L., B\'{e}nisti, D. \& Lefebvre, E.} 2008 Exact
  relativistic kinetic theory of an electron-beam–plasma system: Hierarchy of
  the competing modes in the system-parameter space. {\em Phys. Rev. Lett.\/}
  {\bf 100}, 205008.

\bibitem[Bret {\em et~al.\/}(2010)Bret, Gremillet \& Dieckmann]{BretPoPReview}
{\sc Bret, A., Gremillet, L. \& Dieckmann, M.~E.} 2010 Multidimensional
  electron beam-plasma instabilities in the relativistic regime. {\em Physics
  of Plasmas\/} {\bf 17}, 120501.

\bibitem[Bret {\em et~al.\/}(2013)Bret, Stockem, Fi\'{u}za, Ruyer, Gremillet,
  Narayan \& Silva]{BretPoP2013}
{\sc Bret, A., Stockem, A., Fi\'{u}za, F., Ruyer, C., Gremillet, L., Narayan,
  R. \& Silva, L.~O.} 2013 Collisionless shock formation, spontaneous
  electromagnetic fluctuations, and streaming instabilities. {\em Physics of
  Plasmas\/} {\bf 20}, 042102.

\bibitem[Bret {\em et~al.\/}(2014)Bret, Stockem, Narayan \& Silva]{BretPoP2014}
{\sc Bret, A., Stockem, A., Narayan, R. \& Silva, L.~O.} 2014 Collisionless
  weibel shocks: Full formation mechanism and timing. {\em Physics of
  Plasmas\/} {\bf 21}~(7), 072301.

\bibitem[Davidson {\em et~al.\/}(1972)Davidson, Hammer, Haber \&
  Wagner]{davidsonPIC1972}
{\sc Davidson, Ronald~C., Hammer, David~A., Haber, Irving \& Wagner, Carl~E.}
  1972 Nonlinear development of electromagnetic instabilities in anisotropic
  plasmas. {\em Phys. Fluids\/} {\bf 15}, 317.

\bibitem[Deutsch {\em et~al.\/}(2005)Deutsch, Bret, Firpo \&
  Fromy]{DeutschPRE2005}
{\sc Deutsch, C., Bret, A., Firpo, M.-C. \& Fromy, P.} 2005 Interplay of
  collisions with quasilinear growth rates of relativistic electron-beam-driven
  instabilities in a superdense plasma. {\em Phys. Rev. E\/} {\bf 72}, 026402.

\bibitem[{Dieckmann} {\em et~al.\/}(2014){Dieckmann}, {Sarri}, {Doria}, {Ahmed}
  \& {Borghesi}]{Dieckmann2014NJP}
{\sc {Dieckmann}, M.~E., {Sarri}, G., {Doria}, D., {Ahmed}, H. \& {Borghesi},
  M.} 2014 {Evolution of slow electrostatic shock into a plasma shock mediated
  by electrostatic turbulence}. {\em New Journal of Physics\/} {\bf 16},
  073001.

\bibitem[Fiuza {\em et~al.\/}(2012)Fiuza, Fonseca, Tonge, Mori \&
  Silva]{Fiuza2012}
{\sc Fiuza, F., Fonseca, R.~A., Tonge, J., Mori, W.~B. \& Silva, L.~O.} 2012
  Weibel-instability-mediated collisionless shocks in the laboratory with
  ultraintense lasers. {\em Phys. Rev. Lett.\/} {\bf 108}, 235004.

\bibitem[Godfrey {\em et~al.\/}(1975)Godfrey, Shanahan \& Thode]{Godfrey1975}
{\sc Godfrey, B.~B., Shanahan, W.~R. \& Thode, L.~E.} 1975 Linear theory of a
  cold relativistic beam propagating along an external magnetic field. {\em
  Phys. Fluids\/} {\bf 18}, 346.

\bibitem[{Huntington} {\em et~al.\/}(2015){Huntington}, {Fiuza}, {Ross},
  {Zylstra}, {Drake}, {Froula}, {Gregori}, {Kugland}, {Kuranz}, {Levy}, {Li},
  {Meinecke}, {Morita}, {Petrasso}, {Plechaty}, {Remington}, {Ryutov},
  {Sakawa}, {Spitkovsky}, {Takabe} \& {Park}]{Huntington2015}
{\sc {Huntington}, C.~M., {Fiuza}, F., {Ross}, J.~S., {Zylstra}, A.~B.,
  {Drake}, R.~P., {Froula}, D.~H., {Gregori}, G., {Kugland}, N.~L., {Kuranz},
  C.~C., {Levy}, M.~C., {Li}, C.~K., {Meinecke}, J., {Morita}, T., {Petrasso},
  R., {Plechaty}, C., {Remington}, B.~A., {Ryutov}, D.~D., {Sakawa}, Y.,
  {Spitkovsky}, A., {Takabe}, H. \& {Park}, H.-S.} 2015 {Observation of
  magnetic field generation via the Weibel instability in interpenetrating
  plasma flows}. {\em Nature Physics\/} {\bf 11}, 173--176.

\bibitem[Jackson(1998)]{jackson1998}
{\sc Jackson, J.D.} 1998 {\em Classical Electrodynamics\/}. Wiley.

\bibitem[Lyubarsky \& Eichler(2006)]{lyubarsky06}
{\sc Lyubarsky, Y. \& Eichler, D.} 2006 Are {Gamma}-ray bursts mediated by the
  weibel instability? {\em Astrophys. J.\/} {\bf 647}, 1250.

\bibitem[Marcowith {\em et~al.\/}(2016)Marcowith, Bret, Bykov, Dieckman, Drury,
  Lembège, Lemoine, Morlino, Murphy, Pelletier, Plotnikov, Reville, Riquelme,
  Sironi \& Novo]{Marco2016}
{\sc Marcowith, A, Bret, A, Bykov, A, Dieckman, M~E, Drury, L~O’C, Lembège, B,
  Lemoine, M, Morlino, G, Murphy, G, Pelletier, G, Plotnikov, I, Reville, B,
  Riquelme, M, Sironi, L \& Novo, A~Stockem} 2016 The microphysics of
  collisionless shock waves. {\em Reports on Progress in Physics\/} {\bf 79},
  046901.

\bibitem[Medvedev \& Loeb(1999)]{Medvedev1999}
{\sc Medvedev, M.~V. \& Loeb, A.} 1999 Generation of magnetic fields in the
  relativistic shock of gamma-ray burst sources. {\em Astrophys. J.\/} {\bf
  526}, 697.

\bibitem[Milosavljevic {\em et~al.\/}(2006)Milosavljevic, Nakar \&
  Spitkovsky]{Milosavljevic2006}
{\sc Milosavljevic, M, Nakar, E \& Spitkovsky, A} 2006 Steady state
  electrostatic layers from weibel instability in relativistic collisionless
  shocks. {\em Astrophys. J.\/} {\bf 637}, 765.

\bibitem[Niemiec {\em et~al.\/}(2012)Niemiec, Pohl, Bret \&
  Wieland]{niemiec2012}
{\sc Niemiec, Jacek, Pohl, Martin, Bret, Antoine \& Wieland, Volkmar} 2012
  Nonrelativistic parallel shocks in unmagnetized and weakly magnetized
  plasmas. {\em The Astrophysical Journal\/} {\bf 759}~(1), 73.

\bibitem[Novo {\em et~al.\/}(2015)Novo, Bret, Fonseca \& Silva]{Bret2015ApJL}
{\sc Novo, A.~Stockem, Bret, A., Fonseca, R.~A. \& Silva, L.~O.} 2015 Shock
  formation in electron-ion plasmas: Mechanism and timing. {\em The
  Astrophysical Journal Letters\/} {\bf 803}~(2), L29.

\bibitem[Pathak {\em et~al.\/}(2015)Pathak, Grismayer, Stockem, Fonseca \&
  Silva]{Pathak2015}
{\sc Pathak, V~B, Grismayer, T, Stockem, A, Fonseca, R~A \& Silva, L~O} 2015
  Spatial-temporal evolution of the current filamentation instability. {\em New
  Journal of Physics\/} {\bf 17}~(4), 043049.

\bibitem[Piran(2005)]{Piran2005}
{\sc Piran, Tsvi} 2005 The physics of gamma-ray bursts. {\em Rev. Mod. Phys.\/}
  {\bf 76}, 1143--1210.

\bibitem[{Sagdeev}(1966)]{Sagdeev66}
{\sc {Sagdeev}, R.~Z.} 1966 {Cooperative Phenomena and Shock Waves in
  Collisionless Plasmas}. {\em Reviews of Plasma Physics\/} {\bf 4}, 23.

\bibitem[Sarri {\em et~al.\/}(2011)Sarri, Dieckmann, Kourakis \&
  Borghesi]{Sarri2011}
{\sc Sarri, G., Dieckmann, M.~E., Kourakis, I. \& Borghesi, M.} 2011 Generation
  of a purely electrostatic collisionless shock during the expansion of a dense
  plasma through a rarefied medium. {\em Phys. Rev. Lett.\/} {\bf 107}, 025003.

\bibitem[Schlickeiser \& Shukla(2003)]{Schlickeiser2003}
{\sc Schlickeiser, R. \& Shukla, P.~K.} 2003 Cosmological magnetic field
  generation by the weibel instability. {\em Astrophys. J. Lett.\/} {\bf 599},
  L57.

\bibitem[Silva {\em et~al.\/}(2003)Silva, Fonseca, Tonge, Dawson, Mori \&
  Medvedev]{SilvaApJ}
{\sc Silva, L.~O., Fonseca, R.~A., Tonge, J.~W., Dawson, J.~M., Mori, W.~B. \&
  Medvedev, M.~V.} 2003 Interpenetrating plasma shells: Near-equipartition
  magnetic field generation and nonthermal particle acceleration. {\em
  Astrophys. J.\/} {\bf 596}, L121--L124.

\bibitem[Silva {\em et~al.\/}(2002)Silva, Fonseca, Tonge, Mori \&
  Dawson]{Silva2002}
{\sc Silva, L.~O., Fonseca, R.~A., Tonge, J.~W., Mori, W.~B. \& Dawson, J.~M.}
  2002 On the role of the purely transverse weibel instability in fast ignitor
  scenarios. {\em Physics of Plasmas\/} {\bf 9}, 2458.

\bibitem[{Spitkovsky}(2005)]{Spitkovsky2005}
{\sc {Spitkovsky}, A.} 2005 {Simulations of relativistic collisionless shocks:
  shock structure and particle acceleration}. In {\em Astrophysical Sources of
  High Energy Particles and Radiation\/} (ed. T.~{Bulik}, B.~{Rudak} \&
  G.~{Madejski}), {\em American Institute of Physics Conference Series\/}, vol.
  801, pp. 345--350.

\bibitem[{Stockem} {\em et~al.\/}(2014){Stockem}, {Fiuza}, {Bret}, {Fonseca} \&
  {Silva}]{Stockem2014}
{\sc {Stockem}, A., {Fiuza}, F., {Bret}, A., {Fonseca}, R.~A. \& {Silva},
  L.~O.} 2014 {Exploring the nature of collisionless shocks under laboratory
  conditions}. {\em Scientific Reports\/} {\bf 4}, 3934.

\bibitem[{Stockem} {\em et~al.\/}(2006){Stockem}, {Lerche} \&
  {Schlickeiser}]{Stockem2006ApJ}
{\sc {Stockem}, A., {Lerche}, I. \& {Schlickeiser}, R.} 2006 {On the Physical
  Realization of Two-dimensional Turbulence Fields in Magnetized Interplanetary
  Plasmas}. {\em Astrophys. J.\/} {\bf 651}, 584--589.

\bibitem[{Treumann}(2009)]{Treumann2009}
{\sc {Treumann}, R.~A.} 2009 {Fundamentals of collisionless shocks for
  astrophysical application, 1. Non-relativistic shocks}. {\em Astronomy \&
  Astrophysics Reviews\/} {\bf 17}, 409--535.

\bibitem[{Vietri} {\em et~al.\/}(2003){Vietri}, {De Marco} \&
  {Guetta}]{Vietri2003ApJ}
{\sc {Vietri}, M., {De Marco}, D. \& {Guetta}, D.} 2003 {On the Generation of
  Ultra-High-Energy Cosmic Rays in Gamma-Ray Bursts: A Reappraisal}. {\em
  Astrophysical Journal\/} {\bf 592}, 378--389.

\bibitem[Zel'dovich \& Raizer(2002)]{Zeldovich}
{\sc Zel'dovich, Ya~B \& Raizer, Yu~P} 2002 {\em Physics of shock waves and
  high-temperature hydrodynamic phenomena\/}. Dover Publications.

\end{thebibliography}
\end{document}